\documentclass[a4paper,12pt]{article}
\usepackage{dsfont}
\usepackage{amsmath}
\usepackage{accents}
\usepackage{bm}
\usepackage[english]{babel}
\usepackage{graphicx}

\begin{document}

\centerline {\LARGE{Preparation of an arbitrary two-qubit quantum gate on two spins}}
\centerline {\LARGE{with an anisotropic Heisenberg interaction}}
\medskip
\centerline {A. R. Kuzmak}
\centerline {\small \it E-Mail: andrijkuzmak@gmail.com}
\medskip
\centerline {\small \it Department for Theoretical Physics, Ivan Franko National University of Lviv,}
\medskip
\centerline {\small \it 12 Drahomanov St., Lviv, UA-79005, Ukraine}

{\small

We consider the two-step method [A. R. Kuzmak, V. M. Tkachuk, Phys. Lett. A \textbf{378}, 1469 (2014)]
for preparation of an arbitrary quantum gate on two spins with anisotropic Heisenberg interaction.
At the first step, the system evolves during some period of time. At the second step, we apply pulsed magnetic field individually
to each spin. We obtain the conditions for realization of SWAP, iSWAP, $\sqrt{SWAP}$ and entangled gates.
Finally, we consider the implementation of this method on the physical system of ultracold atoms in optical lattice.

\medskip

}

\section{Introduction\label{sec1}}

Quantum calculations consist of the circuit of unitary operators which provides the quantum evolution of a system
of qubits. This circuit of operators is called quantum algorithm, and a particular operator is called a quantum gate.
An arbitrary gate which provides the transformation of one qubit can be created using two different one-qubit gates (see, for example, \cite{KCQC,KQCBA}).
Also it was shown that any two one-qubit gates and one two-qubit
gate is suffice for the preparation of an arbitrary quantum unitary transformation of $n$ qubits system \cite{BEGQC}.

The problem of implementation of effective computations is related to the problem of the system control. This in turn
requires finding physical systems which allow to prepare the quantum gates. These systems must be isolated from their environment
for providing a high degree of coherence. The states prepared on such systems must be measured with high fidelity. Such systems were
suggested in many papers: spins of electrons and nuclei of atoms \cite{qdots1,qdots2,phosphorus3,phosphorus1}, superconducting circuits
\cite{supcond1,supcond2,supcond3,supcond4}, trapped ions \cite{SchrodCat1,EQSSTI,ITQLLWR,SchrodCat2,ETDIITIQSHI,QSDEGHTI,QSWTI}, and ultracold atoms
\cite{opticallattice1,opticallattice18,opticallattice5,opticallattice19,opticallattice6,opticallattice7,opticallattice9} etc.

Also it is important to find methods which allow to prepare specific quantum gates during the minimal possible time. This fact allows
to provide fast quantum computations and save the energy resources. For instance, the conditions for time-optimal evolution
of a spin-$\frac{1}{2}$ in the magnetic field were obtained in \cite{FSM2,TMTSPMF,FSM,TOSTSS}. Similar problems for the spin-$1$
and arbitrary spin were solved in \cite{QBS1} and \cite{brachass}, respectively. Also it was widely studied the implementation
of quantum gate on two spins with different types of interaction \cite{twosqg1,twosqg2,twosqg3,twosqg4,twosqg5,twosqg6,twosqg7}.

In previous paper \cite{KTTSM} we obtained the two-step method for the preparation of an arbitrary quantum state of two spins
with isotropic Heisenberg interaction. In present paper, we consider the preparation of an arbitrary quantum gate on two spins with an
anisotropic Heisenberg interaction (Sec. \ref{sec2}). Conditions for preparation of SWAP, iSWAP, square root of SWAP and entangled gates
are obtained in Sec. \ref{sec3}. In Sec. \ref{sec4} we consider the implementation of this method on a system of ultracold atoms.
Conclusions are given in Sec. \ref{sec5}.

\section{Method for preparation of an arbitrary two-qubit quantum gate \label{sec2}}

We consider the method which allows to prepare an arbitrary quantum gate on two spins with anisotropic Heisenberg interaction
\begin{eqnarray}
H_{int}=\frac{J}{4}\left(\sigma_x^1\sigma_x^2+\sigma_y^1\sigma_y^2+\gamma \sigma_z^1\sigma_z^2\right),
\label{form1}
\end{eqnarray}
where $\sigma_{\alpha}^i$ are the Pauli matrices for $i$-th spin,
$J$ is the interaction coupling and $\gamma$ is any real number that defines the anisotropy of the system.
When $\gamma=1$ then interaction between two spins is represented by the isotropic Heisenberg Hamiltonian.
In the case of $\gamma=0$ the interaction between two spins is described by the Heisenberg $XX$ model.
Hamiltonian (\ref{form1}) has one two-fold degenerate eigenvalue $\gamma J/4$ with the eigenvectors
$\vert\uparrow\uparrow\rangle$, $\vert\downarrow\downarrow\rangle$,
and eigenvalues $\left(-\gamma J+2J\right)/4$, $\left(-\gamma J-2J\right)/4$ with the eigenvectors
$1/\sqrt{2}\left(\vert\uparrow\downarrow\rangle + \vert\downarrow\uparrow\rangle\right)$,
$1/\sqrt{2}\left(\vert\uparrow\downarrow\rangle - \vert\downarrow\uparrow\rangle\right)$, respectively.

The method which we use is similar to that proposed by us in paper \cite{KTTSM} and consists of two steps. At the first step,
the evolution of two spins is defined by unitary operator with Hamiltonian (\ref{form1})
\begin{eqnarray}
&&U_{int}=e^{-iH_{int}t}=\left[1+\left(\cos\left(\frac{Jt}{2}\right)-1\right)\frac{1}{2}\left(1-\sigma_z^1\sigma_z^2\right)-\frac{i}{2}\sin\left(\frac{Jt}{2}\right)\left(\sigma_x^1\sigma_x^2+\sigma_y^1\sigma_y^2\right)\right]\nonumber\\
&&\times \left[\cos\left(\frac{\gamma Jt}{4}\right)-i\sin\left(\frac{\gamma Jt}{4}\right) \sigma_z^1\sigma_z^2\right].
\label{form2}
\end{eqnarray}
Here we use the fact that $XX$ and $ZZ$ parts of Hamiltonian mutually commute and that
\begin{eqnarray}
\left(\sigma_x^1\sigma_x^2+\sigma_y^1\sigma_y^2\right)^{2}=2\left(1-\sigma_z^1\sigma_z^2\right),\quad \left(\sigma_z^1\sigma_z^2\right)^2=1.\nonumber
\end{eqnarray}
We set $\hbar=1$, which means that the
energy is measured in the frequency units. In the basis labelled by
$\vert\uparrow\uparrow\rangle$, $\vert\uparrow\downarrow\rangle$,
$\vert\downarrow\uparrow\rangle$ and
$\vert\downarrow\downarrow\rangle$, operator (\ref{form2}) can be represented as
\begin{eqnarray}
U_{int}=\left( \begin{array}{ccccc}
e^{-i\frac{\gamma Jt}{4}} & 0 & 0 & 0\\
0 & \cos\left(\frac{Jt}{2}\right)e^{i\frac{\gamma Jt}{4}} & -i\sin\left(\frac{Jt}{2}\right)e^{i\frac{\gamma Jt}{4}} & 0 \\
0 & -i\sin\left(\frac{Jt}{2}\right)e^{i\frac{\gamma Jt}{4}} & \cos\left(\frac{Jt}{2}\right)e^{i\frac{\gamma Jt}{4}} & 0 \\
0 & 0 & 0 & e^{-i\frac{\gamma Jt}{4}}
\end{array}\right).
\label{form3}
\end{eqnarray}
At the second step, at the moment of time $t_1$ we apply pulsed magnetic fields individually to each spin
\begin{eqnarray}
H_{mf}=\frac{\omega_1}{2}\mbox{\boldmath{$ \sigma$}}^1 \cdot {\bf n}^1\delta(t-t_1)+\frac{\omega_2}{2}\mbox{\boldmath{$ \sigma$}}^2 \cdot {\bf n}^2\delta(t-t_1),
\label{form4}
\end{eqnarray}
where $\omega_i$ is proportional to the strength of the magnetic field for $i$-th spin,
$\bf{n}^{\it i}=\left(\sin\theta_{\it i}\cos\phi_{\it i},\sin\theta_{\it i}\sin\phi_{\it i},\cos\theta_{\it i}\right)$ is a unit vector
defined by the spherical angles $\theta_i$, $\phi_i$ which determine the direction of the magnetic field for the $i$-th spin,
and $\delta(t-t_1)$ is Dirac's delta function which allows us to neglect the interaction between spins when the magnetic field is applied.

So, the operator of evolution for two spins in the magnetic fields at the moment of time $t_1$ takes the form
\begin{eqnarray}
&&U_{mf}=e^{-i\left(\frac{\omega_1}{2}\bm{\sigma}^1 \cdot {\bf n}^1 +\frac{\omega_2}{2}\bm{\sigma}^2 \cdot {\bf n}^2\right)}\nonumber\\
&&=\left(\cos\frac{\omega_1}{2}-i\mbox{\boldmath{$ \sigma$}}^1 \cdot {\bf n}^1\sin\frac{\omega_1}{2} \right)\left(\cos\frac{\omega_2}{2}-i\mbox{\boldmath{$ \sigma$}}^2 \cdot {\bf n}^2\sin\frac{\omega_2}{2} \right).
\label{form5}
\end{eqnarray}
Here we use the fact that $\mbox{\boldmath{$ \sigma$}}^1 \cdot {\bf n}^1$ and $\mbox{\boldmath{$ \sigma$}}^2 \cdot {\bf n}^2$ mutually commute and $\left(\mbox{\boldmath{$ \sigma$}}^i \cdot {\bf n}^i\right)^2=1$.
The first and the second factors in this operator describe quantum evolution of the first and second spins under the external magnetic fields,
respectively. In the matrix representation operator (\ref{form5}) has the form
\begin{eqnarray}
U_{mf}=U_1U_2,
\label{form6}
\end{eqnarray}
where
\begin{eqnarray}
{\scriptsize
U_1=\left( \begin{array}{ccccc}
\cos\frac{\omega_1}{2}-i\sin\frac{\omega_1}{2}\cos\theta_1 & 0 & -i\sin\frac{\omega_1}{2}\sin\theta_1 e^{-i\phi_1} & 0\\
0 & \cos\frac{\omega_1}{2}-i\sin\frac{\omega_1}{2}\cos\theta_1 & 0 & -i\sin\frac{\omega_1}{2}\sin\theta_1e^{-i\phi_1} \\
-i\sin\frac{\omega_1}{2}\sin\theta_1e^{i\phi_1} & 0 & \cos\frac{\omega_1}{2}+i\sin\frac{\omega_1}{2}\cos\theta_1  & 0 \\
0 & -i\sin\frac{\omega_1}{2}\sin\theta_1e^{i\phi_1} & 0 & \cos\frac{\omega_1}{2}+i\sin\frac{\omega_1}{2}\cos\theta_1
\end{array}\right)}
\label{form6_1}
\end{eqnarray}
and
\begin{eqnarray}
{\scriptsize
U_2=\left( \begin{array}{ccccc}
\cos\frac{\omega_2}{2}-i\sin\frac{\omega_2}{2}\cos\theta_2 & -i\sin\frac{\omega_2}{2}\sin\theta_2 e^{-i\phi_2} & 0  & 0\\
-i\sin\frac{\omega_2}{2}\sin\theta_2 e^{i\phi_2} & \cos\frac{\omega_2}{2}+i\sin\frac{\omega_2}{2}\cos\theta_2 & 0 & 0 \\
0 & 0 & \cos\frac{\omega_2}{2}-i\sin\frac{\omega_2}{2}\cos\theta_2  & -i\sin\frac{\omega_2}{2}\sin\theta_2 e^{-i\phi_2} \\
0 & 0 & -i\sin\frac{\omega_2}{2}\sin\theta_2 e^{i\phi_2} & \cos\frac{\omega_2}{2}+i\sin\frac{\omega_2}{2}\cos\theta_2
\end{array}\right)}.
\label{form6_2}
\end{eqnarray}

\section{Preparation of quantum gates \label{sec3}}

Using the method from previous section let us obtain the conditions for realization of some two-qubit quantum gate.
The target gate $W$ is achieved from the circuit of unitary operators with modulo a global phase as follows
\begin{eqnarray}
W=e^{i\chi}U(\lambda_1)U(\lambda_2)\ldots ,
\label{form7}
\end{eqnarray}
where $\lambda_i$ is the set of parameters which allows to achieve the target gate and $\chi$ is some phase. In our case
the circuit consists of two unitary operators, namely, operator which provide interaction between spins (\ref{form2})
and action of external magnetic fields (\ref{form5}). To obtain the parameters for realization some quantum gate we consider
the fidelity between circuit of unitary operators and target operator \cite{FIDELITY}
\begin{eqnarray}
F=\frac{1}{4}\Re(\textrm{Tr}[W^+U_{mf}U_{int}]).
\label{form8}
\end{eqnarray}
So, we obtain the target gate when the fidelity reaches the value $F=1$.

Let us demonstrate this explicitly on some examples. Firstly, we consider the implementation of the SWAP and iSWAP gates
\begin{eqnarray}
W_{SWAP_{\beta}}=\left( \begin{array}{ccccc}
1 & 0 & 0  & 0\\
0 & 0 & e^{i\beta} & 0 \\
0 & e^{i\beta} & 0  & 0 \\
0 & 0 & 0 & 1
\end{array}\right),
\label{form7_1}
\end{eqnarray}
where $\beta$ takes two values $0$ and $\pi/2$ for SWAP and iSWAP gates, respectively. These gates exchange the states of two qubits.
Using equation (\ref{form8}) with (\ref{form3}), (\ref{form6}), and (\ref{form7_1}) we obtain
a set of conditions for realization SWAP and iSWAP gates. These gates we achieve for $Jt=\pi+2\pi n$, where $n\in \mathds{Z}$.
Then $\chi=\gamma\left(\pi/4+\pi n/2\right)$. So, the SWAP operator additionally satisfies the following sets of conditions:\\
1) for $\gamma=4p+1$ the magnetic fields should be switched off ($\omega_1=\omega_2=0$)\\ and\\ 2) for $\gamma=4p+3$
the magnetic fields must be satisfied the following conditions $\omega_1=-\omega_2=\pi$ and $\theta_1=\theta_2=0$,
where $p\in \mathds{Z}$.\\ In the first case we just allow the system evolves during the time $t_1=\vert\left(\pi+2\pi n\right)/J\vert$.
In the second case the system evolves during the time $t_1=\vert\left(\pi+2\pi n\right)/J\vert$, and at the moment of time $t_1$ we apply to each
spin along the $z$-axis opposite magnetic fields with the same magnitude. The iSWAP operator can be implemented for the following cases:\\
1) $\gamma=4p$, then for even $n$ the parameters which determine the magnetic field take the form $\omega_1=-\omega_2=\pi$ and
$\theta_1=\theta_2=0$, and for odd $n$ the magnetic fields should be switched off;\\
2) $\gamma=4p+2$, for even $n$ the magnetic fields should be switched off, and for odd $n$ the magnetic fields satisfy the
following conditions $\omega_1=-\omega_2=\pi$ and $\theta_1=\theta_2=0$.

The next gate which we study is the square root of SWAP operator
\begin{eqnarray}
W_{\sqrt{SWAP}}=\left( \begin{array}{ccccc}
1 & 0 & 0  & 0\\
0 & \frac{1}{2}(1+i) & \frac{1}{2}(1-i) & 0 \\
0 & \frac{1}{2}(1-i) & \frac{1}{2}(1+i) & 0 \\
0 & 0 & 0 & 1
\end{array}\right).
\label{form7_2}
\end{eqnarray}
This gate is a fundamental operator because many unitary operations can be implemented by the square root of SWAP and single qubit
operations \cite{twosqg7}. This gate can be achieved on two spins for $Jt=\pi/2+2\pi n$ and $\gamma=4p+1$, where
$n,p\in \mathds{Z}$. Then $\omega_1=\omega_2=0$ if $p$ is even and $\omega_1=-\omega_2=\pi$, $\theta_1=\theta_2=0$ if $p$ is odd.
Here $\chi=\gamma\left(\pi/8+\pi n/2\right)$.

Also we can prepare the quantum gate which allows to achieve the entangled states. For this purpose $Jt$ must be equal $\pi/2+\pi n$,
where $n\in \mathds{Z}$. The easiest way to obtain the gate which creates the following Bell states
$\vert\Psi^{\pm}\rangle=\frac{1}{\sqrt{2}}\left(\vert\uparrow\downarrow\rangle\pm\vert\downarrow\uparrow\rangle\right)$ from
the states $\vert\uparrow\downarrow\rangle$ and $\vert\downarrow\uparrow\rangle$ is to put $Jt=\pi/2$, $\gamma=0$,
$\omega_1-\omega_2=\pi/2$ and $\theta_1=\theta_2=0$. This gate has the form
\begin{eqnarray}
W_{ENT}=\left( \begin{array}{ccccc}
e^{-\frac{i}{2}\left(\omega_1+\omega_2\right)} & 0 & 0  & 0\\
0 & \frac{1}{\sqrt{2}}e^{-i\frac{\pi}{4}} & \frac{1}{\sqrt{2}}e^{-i\frac{3\pi}{4}} & 0 \\
0 & \frac{1}{\sqrt{2}}e^{-i\frac{\pi}{4}} & \frac{1}{\sqrt{2}}e^{i\frac{\pi}{4}} & 0 \\
0 & 0 & 0 & e^{\frac{i}{2}\left(\omega_1+\omega_2\right)}
\end{array}\right).
\label{form7_3}
\end{eqnarray}

The similar gate which allows to create the following pair of Bell states
$\vert\Phi^{\pm}\rangle=\frac{1}{\sqrt{2}}\left(\vert\uparrow\uparrow\rangle\pm\vert\downarrow\downarrow\rangle\right)$
can be achieved from gate (\ref{form7_3}) as follows
\begin{eqnarray}
e^{-i\frac{\pi}{2}\sigma_x^1}W_{ENT}e^{i\frac{\pi}{2}\sigma_x^1}=\left( \begin{array}{ccccc}
\frac{1}{\sqrt{2}}e^{i\frac{\pi}{4}} & 0 & 0  & \frac{1}{\sqrt{2}}e^{-i\frac{\pi}{4}}\\
0 & e^{\frac{i}{2}\left(\omega_1+\omega_2\right)}& 0 & 0 \\
0 & 0 & e^{-\frac{i}{2}\left(\omega_1+\omega_2\right)}& 0 \\
\frac{1}{\sqrt{2}}e^{-i\frac{3\pi}{4}} & 0 & 0 & \frac{1}{\sqrt{2}}e^{-i\frac{\pi}{4}}
\end{array}\right).
\label{form7_4}
\end{eqnarray}
This gate allows to prepare $\vert\Phi^+\rangle$ and $\vert\Phi^-\rangle$ states from $\vert\downarrow\downarrow\rangle$
and $\vert\uparrow\uparrow\rangle$ states, respectively.

\section{Physical implementation on ultracold atoms \label{sec4}}

In this section we propose the implementation of two-qubit quantum gates on physical system of ultracold atoms
in optical lattice. The ultracold atoms are very useful for realization of quantum calculations
\cite{opticallattice6,opticallattice7,opticallattice9,opticallattice8,opticallattice10,opticallattice11,opticallattice12,opticallattice13,opticallattice20} because such systems allow
easily simulating an effective spin system with controlled interaction
\cite{opticallattice1,opticallattice5,opticallattice19,opticallattice14,opticallattice15,opticallattice16,opticallattice17,opticallattice2,opticallattice3,opticallattice4}.
To prepare an effective spin system the atoms with two relevant internal states, which define the effective spin projection
$\sigma=\uparrow,\downarrow$, are used. Using periodic potential $V_{\sigma}\sin^2\left({\bf k}{\bf r}\right)$
generated by the laser beams with wave vectors ${\bf k}$ a set of $N$ ultracold bosonic or fermionic atoms
are trapped in optical lattice with certain directions
\cite{opticallattice1,opticallattice5,opticallattice0,opticallattice2,opticallattice3,opticallattice4,opticallattice21}.
The Hamiltonian of the two trapped atoms is given by
\begin{eqnarray}
H=-\sum_{\sigma}\left(t_{\sigma}a^+_{1\sigma}a_{2\sigma} + H.c.\right)+\frac{1}{2}\sum_{i\sigma}U_{\sigma\sigma}n_{i\sigma}\left(n_{i\sigma}-1\right)+U_{\uparrow\downarrow}\sum_{i}n_{i\uparrow}n_{i\downarrow},
\label{form9}
\end{eqnarray}
where $i$ labels the site, $a_{i\sigma}$ are the bosonic (or fermionic) annihilation operators, $n_{i\sigma}=a^+_{i\sigma}a_{i\sigma}$.
The tunnelling and one-site interaction energies for the cubic lattice are defined by \cite{opticallattice1,opticallattice18,opticallattice5}
\begin{eqnarray}
&&t_{\sigma}=\frac{4}{\sqrt{\pi}}E_r\left(\frac{V_{\sigma}}{E_r}\right)^{3/4}\exp{\left[-2\left(\frac{V_{\sigma}}{E_r}\right)^{1/2}\right]},\nonumber\\
&&U_{\uparrow\downarrow}=\sqrt{\frac{8}{\pi}}ka_{\uparrow\downarrow}E_r\left(\frac{\overline{V}_{\uparrow\downarrow}}{E_r}\right)^{3/4},\nonumber\\
&&U_{\sigma\sigma}^{b}=\sqrt{\frac{8}{\pi}}ka_{\sigma\sigma}E_r\left(\frac{V_{\sigma}}{E_r}\right)^{3/4},\quad U_{\sigma\sigma}^{f}=2E_r\sqrt{\frac{V_{\sigma}}{E_r}},
\label{form10}
\end{eqnarray}
where $E_r=\hbar^2k^2/(2m)$ is the atomic recoil energy, $a_{\sigma\sigma}$ is the scattering length between two atoms
and $\overline{V}_{\uparrow\downarrow}=4V_{\uparrow}V_{\downarrow}/(V_{\uparrow}^{1/2}+V_{\downarrow}^{1/2})^2$
is the spin average potential in each direction and $U_{\sigma\sigma}^{b,f}$ correspond to bosonic and fermionic atoms, respectively.

So, in the regime where $t_{\sigma}\ll U_{\sigma\sigma},\ U_{\uparrow\downarrow}$ the parameter $t_{\sigma}$ can be considered
as a perturbation parameter. Then Hamiltonian (\ref{form9}) with respect to the second-order of perturbation theory is
equivalent to Hamiltonian (\ref{form1}). Then,
\begin{eqnarray}
J=\pm \frac{t_{\uparrow}t_{\downarrow}}{U_{\uparrow\downarrow}},\quad
\gamma J=\frac{t_{\uparrow}^2+t_{\downarrow}^2}{2U_{\uparrow\downarrow}}-\frac{t_{\uparrow}^2}{U_{\uparrow\uparrow}}-\frac{t_{\downarrow}^2}{U_{\downarrow\downarrow}},
\label{form11}
\end{eqnarray}
where the $"+"$ and $"-"$ signs in $J$ correspond to the cases of fermionic and bosonic atoms, respectively, and for fermionic atoms
the last two terms in $\gamma J$ vanish since $U_{\sigma\sigma}\gg U_{\uparrow\downarrow}$. So, as we can see from (\ref{form10}),
changing intensity of the beams we can control the interaction between the effective spins.

The single-spin rotation in optical lattices can be provided using method described in paper \cite{opticallattice17}. This
method is based on the interference of laser beams. This experimental techniques allows appling
the effective magnetic field with the value defined by the Rabi frequency $\Omega$ individually to each spin. The time of single-spin rotation
on angle $\eta$ around some axis depends on $\Omega$ as follows $t_{rot}=\eta/\Omega$. So, as we can see the grater $\Omega$ the shorter time for
rotation is needed. Therefore, the preparation of the effective local pulse magnetic field, which is necessary for implementation of considered methods
in Section \ref{sec2}, can be experimentally realized due to very rapidly rotating of a single spin. For this purpose the following condition
should be satisfied $\Omega\gg J$.

For instance, let us find the conditions on Hamiltonian (\ref{form9}) for the preparation of $SWAP$ (\ref{form7_1})
and $\sqrt{SWAP}$ (\ref{form7_2}) gates. The simplest way for implementation of this gates requires the preparation
of the isotropic Haisenberg Hamiltonian. From expression (\ref{form11}) it follows that this Hamiltonian can be created
when the following conditions are satisfied: $t_{\uparrow}=t_{\downarrow}$ and $U_{\uparrow\downarrow}=U_{\uparrow\uparrow}=U_{\downarrow\downarrow}$.
The implementation of $iSWAP$ gate requres the preparation at least the $XX$ Hamiltonian. Similarly as in the previous case we obtain the conditions
$2U_{\uparrow\downarrow}=U_{\uparrow\uparrow}=U_{\downarrow\downarrow}$ which should be imposed on the system of ultracold atoms.
Another example to be addressed in regards the implementation of entangled gate (\ref{form7_3}).
This gate also requires the preparetion of the $XX$ Hamiltonian.

As noted in the introduction, to prepare the quantum gate with high fidelity the coherence time $t_c$ of the system should be longer than
the time of preparation of a quantum gate $t_{gate}$ ($t_{gate}\leq t_c$). So, we can obtain the limitation on the interaction caupling $J$
to achieve the maximally possible fidelity for the preparetion of a particular quantum gate. Using the results from the Section \ref{sec3}
for minimally possible time of the implementation of quantum gates ($n=0$) let us find this limit for each quantum gate. To implement
the $SWAP_{\beta}$ gate (\ref{form7_1}) the following condition $J\geq \pi/t_c$ should be satiesfied. In the case of $\sqrt{SWAP}$ gate (\ref{form7_2})
and entangled gates (\ref{form7_3}), (\ref{form7_4}) we obtain the following condition $J\geq \pi/(2t_c)$. So, the interaction coupling is limited
by the coherence time. The longer the coherence time the smaller the interaction coupling can be chosen and vice versa. However, on the other hand,
as we know from the present section, to prepare effective Heisenberg Hamiltonian (\ref{form1}) on the ultracold atoms the tunneling parameters
$t_{\sigma}$ must be much smaller than one-site interaction energies $U_{\sigma\sigma}$ and $U_{\uparrow\downarrow}$. This fact makes the interaction
coupling a small value, that in turn requires a sufficiently long coherence time of system. Therefore, the implementation of the quantum calculations on
ultracold atoms requires the system with a sufficiently long coherence time. One such system is the $^{87}$Rb atoms in optical lattice
\cite{opticallattice20,opticallattice2}. Let us obtain the limitation on $J$ for preparation of quantum gates considered in Section \ref{sec3}
on this system.

In paper \cite{opticallattice20} it was prepared the array of pairs of $^{87}$Rb atoms in a three-dimensional optical lattice. The interaction
between each pair is described by Heisenberg Hamiltonian (\ref{form1}). The coherence time achieved by the authors for such pair is more than 10 ms.
So, to implement the quantum gates with maximal fidelity the interaction coupling $J$ should not be less than $\approx 0,314$kHz
for $SWAP_{\beta}$ gate (\ref{form7_1}) and $\approx 0,157$kHz for $\sqrt{SWAP}$ (\ref{form7_2}) and entangled gates (\ref{form7_3}), (\ref{form7_4}).
Also it is worth to noting that the authors of paper \cite{opticallattice20} realized on such system the $SWAP$ and $\sqrt{SWAP}$ gates.

\section{Conclusion \label{sec5}}

We considered the method which allows to prepare an arbitrary quantum gate on two spins defined
by anisotropic Heisenberg interaction (\ref{form1}). This method consists of two steps. At the first step, the system evolves during
some period of time $t_1$. At the second step, at the moment of time $t_1$ we apply pulsed magnetic field individually to each spin.
Choosing the parameter of anisotropy, time of evolution, values and directions of magnetic fields we can prepare the required quantum
gate. We obtained the conditions for preparation of SWAP, iSWAP, square root of SWAP and entangled gates. Finally we considered the implementation
of this method on physical system of ultracold atoms in optical lattice. This system is very useful for realization of quantum gates
because it allows easily simulating an effective spin system with controlled interaction. We obtain the conditions
for realization of some quantum gates and limitation on the value of interaction coupling between spins.

\section{Acknowledgement}

The author thanks Prof. Andrij Rovenchak and Dr. Taras Verkholyak for useful comments. This work was supported by Project FF-30F (No. 0116U001539)
from the Ministry of Education and Science of Ukraine.

\end{document}